\documentclass[reprint,aps,prL,twocolumn,superscriptaddress]{revtex4-1}

\usepackage{graphicx}
\usepackage{dcolumn}
\usepackage{bm}
\usepackage{amsmath}
\usepackage{braket}
\usepackage[english]{babel}
\usepackage[utf8]{inputenc}
\usepackage[dvipsnames]{xcolor}
\usepackage{soul}
\usepackage{siunitx}
\begin{document}

\title{High-fidelity two-qubit gates in silicon above one Kelvin}

\author{L.~Petit}
\email{l.petit@tudelft.nl}
\affiliation{QuTech and Kavli Institute of Nanoscience, Delft University of Technology, PO Box 5046, 2600 GA Delft, The Netherlands}
\author{M.~Russ}
\affiliation{QuTech and Kavli Institute of Nanoscience, Delft University of Technology, PO Box 5046, 2600 GA Delft, The Netherlands}
\author{H.~G.~J.~Eenink}
\affiliation{QuTech and Kavli Institute of Nanoscience, Delft University of Technology, PO Box 5046, 2600 GA Delft, The Netherlands}
\author{W.~I.~L.~Lawrie}
\affiliation{QuTech and Kavli Institute of Nanoscience, Delft University of Technology, PO Box 5046, 2600 GA Delft, The Netherlands}
\author{J.~S.~Clarke}
\affiliation{Components Research, Intel Corporation, 2501 NE Century Blvd, Hillsboro, Oregon 97124, USA}
\author{L.~M.~K.~Vandersypen}
\affiliation{QuTech and Kavli Institute of Nanoscience, Delft University of Technology, PO Box 5046, 2600 GA Delft, The Netherlands}
\author{M.~Veldhorst}
\email{m.veldhorst@tudelft.nl}
\affiliation{QuTech and Kavli Institute of Nanoscience, Delft University of Technology, PO Box 5046, 2600 GA Delft, The Netherlands}

\pacs{}

\maketitle
\textbf{
Spin qubits in quantum dots define an attractive platform for scalable quantum information because of their compatibility with semiconductor manufacturing \cite{loss_quantum_1998, vandersypen_interfacing_2017}, their long coherence times \cite{veldhorst_addressable_2014}, and the ability to operate at temperatures exceeding one Kelvin \cite{yang2020operation,petit2020universal}. Qubit logic can be implemented by pulsing the exchange interaction \cite{petta_coherent_2005, veldhorst_two-qubit_2015, watson_programmable_2018} or via driven rotations \cite{koppens2006driven, zajac_resonantly_2018, huang2019fidelity, hendrickx2020fast}. Here, we show that these approaches can be combined to execute a multitude of native two-qubit gates in a single device, reducing the operation overhead to perform quantum algorithms. We demonstrate, at a temperature above one Kelvin, single-qubit rotations together with the two-qubit gates CROT, CPHASE and SWAP. Furthermore we realize adiabatic, diabatic and composite sequences to optimize the qubit control fidelity and the gate time. We find two-qubit gates that can be executed within 67 ns and by theoretically analyzing the experimental noise sources we predict fidelities exceeding 99\%. This promises fault-tolerant operation using quantum hardware that can be embedded with classical electronics for quantum integrated circuits.}

Two-qubit gates are at the heart of quantum information science, as they may be used to create entangled states with a complexity beyond what is classically simulatable \cite{arute2019quantum}, and ultimately may enable the execution of practically relevant quantum algorithms \cite{reiher2017elucidating}. Optimizing two-qubit gates is therefore a central aspect across all qubit platforms \cite{ladd2010quantum}. In quantum dot systems, two-qubit gates can be naturally implemented using the exchange interaction between spin qubits in neighbouring quantum dots \cite{loss_quantum_1998}. Pulsing the interaction drives SWAP oscillations when the exchange energy is much larger than the Zeeman energy difference of the qubits \cite{loss_quantum_1998, petta_coherent_2005}, while it results in CPHASE oscillations when the Zeeman energy difference is much larger than the exchange energy \cite{meunier2011efficient}. Single-qubit gates need also to be implemented to access the full two-qubit Hilbert space, and this requires distinguishability between the qubits. This is commonly obtained through the spin-orbit coupling \cite{veldhorst_addressable_2014} or by integrating nanomagnets \cite{kawakami_electrical_2014, yoneda2018quantum}, causing significant Zeeman energy differences. Realizing a high-fidelity SWAP-gate in this scenario would require extremely large values of exchange interaction. For this reason, the CPHASE operation has been the native gate in experimental demonstrations of two-qubit logic when the exchange interaction is pulsed~\cite{veldhorst_two-qubit_2015, watson_programmable_2018}. An alternative implementation of two-qubit logic can be realized by driven rotations, which become state dependent in the presence of exchange interaction. This has been used to realize CROT~\cite{zajac_resonantly_2018, huang2019fidelity, hendrickx2020fast, petit2020universal} and resonant SWAP gates~\cite{sigillito2019coherent}.

While universal quantum logic can be obtained by combinations of single- and two-qubit operations \cite{barenco1995elementary}, the ability to directly execute a multitude of two-qubit gates would reduce the number of operations required to execute quantum algorithms. Here, we take this step and investigate on the same device the implementation of the CROT, SWAP, and CPHASE, which are all essential gates in applications ranging from quantum error correction to long-distance qubit connectivity. We furthermore focus on the optimal implementation of these two-qubit gates and find that in particular the CPHASE and the SWAP can be executed with high-fidelity and in short time scales. Moreover, we demonstrate these operations at temperatures exceeding one Kelvin. The cooling power at these elevated temperatures is much larger and thereby more compatible with the operation of classical electronics, such that quantum integrated circuits based on standard semiconductor technology become feasible \cite{veldhorst_silicon_2017, vandersypen_interfacing_2017, li2018crossbar}.

\begin{figure*}[!t]
	\includegraphics[width=\linewidth]{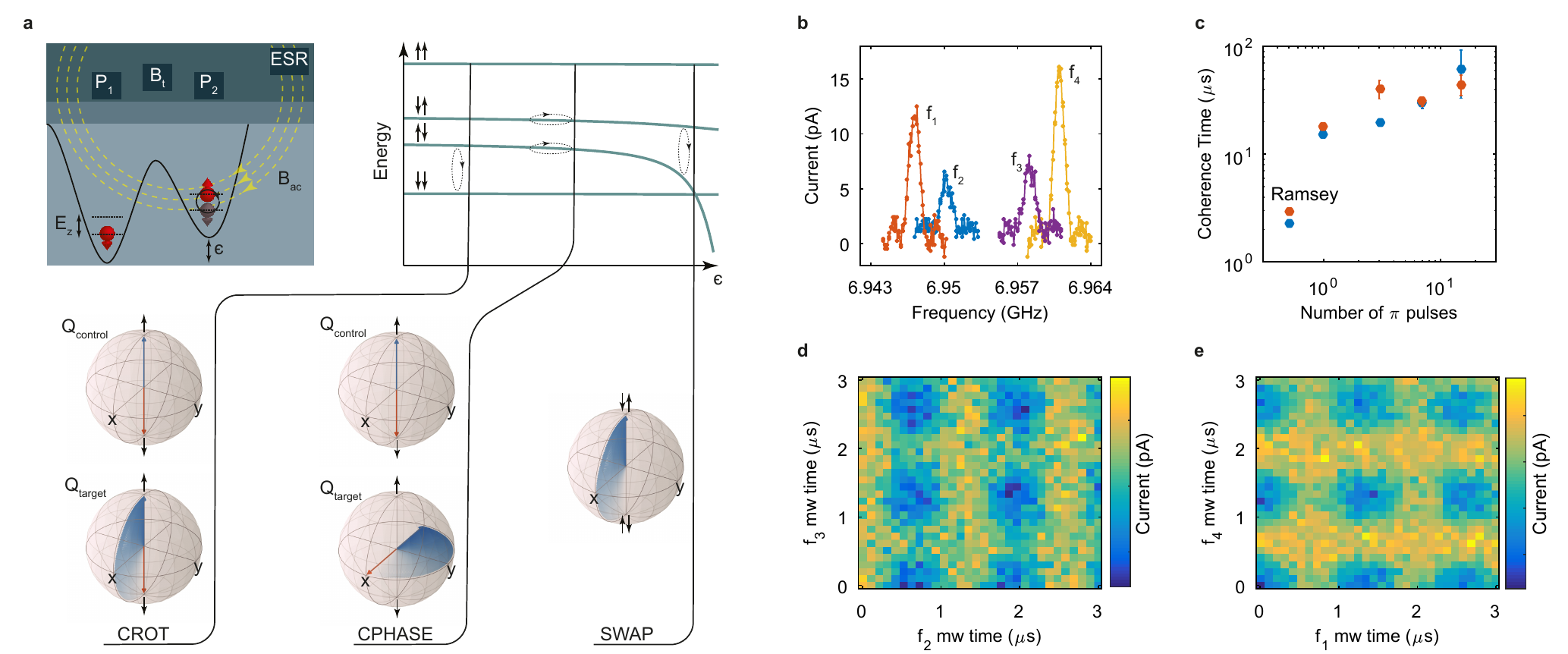}%
	\caption{\textbf{Two-qubit gates and quantum coherence of silicon spin qubits operated at a T = 1.05 K.} 
    	\textbf{a} Schematic representation of the double quantum dot system. Two plunger gates ($P_\mathrm{1}$ and $P_\mathrm{2}$) and one barrier gate ($B_\mathrm{t}$) are used to control the detuning energy $\epsilon$ and the tunnel coupling $t$ between the quantum dots. Spin manipulation occurs via electron-spin-resonance (ESR) using an on-chip microwave line. The energy diagram displays the four electron spin states as a function of $\epsilon$. We exploit both driven rotations and pulsed exchange for coherent control. Controlled rotations (CROTs) can in principle be executed at all points where $J \neq 0$, given that gate times are appropriately set. CPHASE gates are conveniently executed when the exchange interaction is much smaller than the Zeeman energy difference between the qubits, while SWAP oscillations can be realized when the exchange interaction is much larger.	\textbf{b} Using ESR control we find the four resonance frequencies of the two-qubit system. Here, the exchange interaction is tuned to 3 MHz. 
    	\textbf{c} Coherence times as a function of the number of refocusing $\mathrm{\pi}$ pulses. Here, the exchange is set to 2 MHz. The plot includes the dephasing times measured through a Ramsey experiment to allow comparison.
    	\textbf{d-e} Realization of CROT operations. Rabi oscillations of the target qubit are controlled by the spin state of the control qubit. We find controlled rotations on all the four resonance frequencies $f_1, f_2, f_3, f_4$. 
	}
\label{fig:CROT}
\end{figure*}

The experimental two-qubit system is based on electron spin states confined in a silicon double quantum dot as schematically shown in Fig. \ref{fig:CROT}a. The silicon double quantum dot is fabricated using an overlapping  gate architecture on a silicon wafer with an isotopically enriched \textsuperscript{28}Si epilayer of 800 ppm residual concentration of \textsuperscript{29}Si \cite{petit2020universal, lawrie2020quantum}. Qubits $\mathrm{Q1}$ and $\mathrm{Q2}$ are defined with $N_\mathrm{Q1}$ = 5 and $N_\mathrm{Q2}$ = 1, where $N$ is the charge occupancy. Spin readout is performed at the (1,5)-(2,4) charge anticrossing, where the $\ket{\downarrow\uparrow}$ tunnels to the singlet (2,4) charge state, while the other spin states are blocked because of the Pauli exclusion principle. By using an adiabatic pulse from the (2,4) to the (1,5) region, we initialize the system in the $\ket{\downarrow\uparrow}$ state. Because of the limited sensitivity of the single-electron-transistor (SET) that we use for charge readout, we average the single-shot readout traces and subtract a reference signal. We therefore obtain a current signal, proportional to the probability to have a blocked state. We note that the readout fidelity can be further improved, even at these higher temperatures \cite{urdampilleta2018gate}, but here we focus on the coherent control. We perform spin manipulation via electron spin resonance (ESR) using an on-chip aluminum microwave antenna. All measurements have been performed in a dilution refrigerator at a temperature of $T_\mathrm{fridge}=\SI{1.05}{\kelvin}$ and with an external magnetic field of $B_\mathrm{ext}=\SI{250}{\milli\tesla}$. 

We control the exchange interaction $J$ via the detuning $\epsilon$ between the two quantum dots and we measure couplings from $J=\SI{2}{\mega\hertz}$ up to $J=\SI{45}{\mega\hertz}$, as shown in Supp. Fig. 1a. By fitting the exchange spectrum we extract a Zeeman energy difference between the two qubits $\Delta E_\mathrm{z}=\SI{11}{\mega\hertz}$. The fitting suggests a negligible dependence of $\Delta E_\mathrm{z}$ on detuning, further supported by the small magnetic field applied and the absence of external magnetic gradients. Figure \ref{fig:CROT}b shows the four resonance frequencies of the two-qubit system when $J=\SI{3}{\mega\hertz}$. At this value of exchange interaction we tune the $\mathrm{\pi}$-rotation times to be $t_\mathrm{CROT}=\SI{660}{\nano\second}$ such that we synchronize the Rabi oscillations of the target transition with the closest off-resonant transition in order to suppress crosstalk~\cite{russ2018high}. From Ramsey experiments on frequencies $f_1$ and $f_4$ we measure dephasing times $T^*_{2,\mathrm{Q1}}=\SI{2.3}{\micro\second}$ and $T^*_{2,\mathrm{Q2}} = \SI{2.9}{\micro\second}$. The Carr-Purcell-Meiboom-Gill (CPMG) pulse sequence can extend the coherence times, by filtering out the low frequency noise. As shown in Fig. \ref{fig:CROT}c, we measure a maximum $T_{2,\mathrm{Q1}}=\SI{63}{\micro\second}$ and $T_{2,\mathrm{Q2}} =\SI{44}{\micro\second}$ when 15 refocusing pulses are applied, setting new benchmarks for the coherence time of quantum dot spin qubits at temperatures above one Kelvin. 

When the exchange interaction is set to a non-zero value, it is possible to realize the CROT via driven rotations since the resonance frequency of one qubit depends on the state of the other qubit. This CROT gate is a universal two-qubit gate and equivalent to a CNOT gate up to single qubit phases \cite{petit2020universal}. Figures \ref{fig:CROT}d-e show controlled rotations by setting both configurations of target and control qubits.

\begin{figure*}[!t]
	\includegraphics[width=\linewidth]{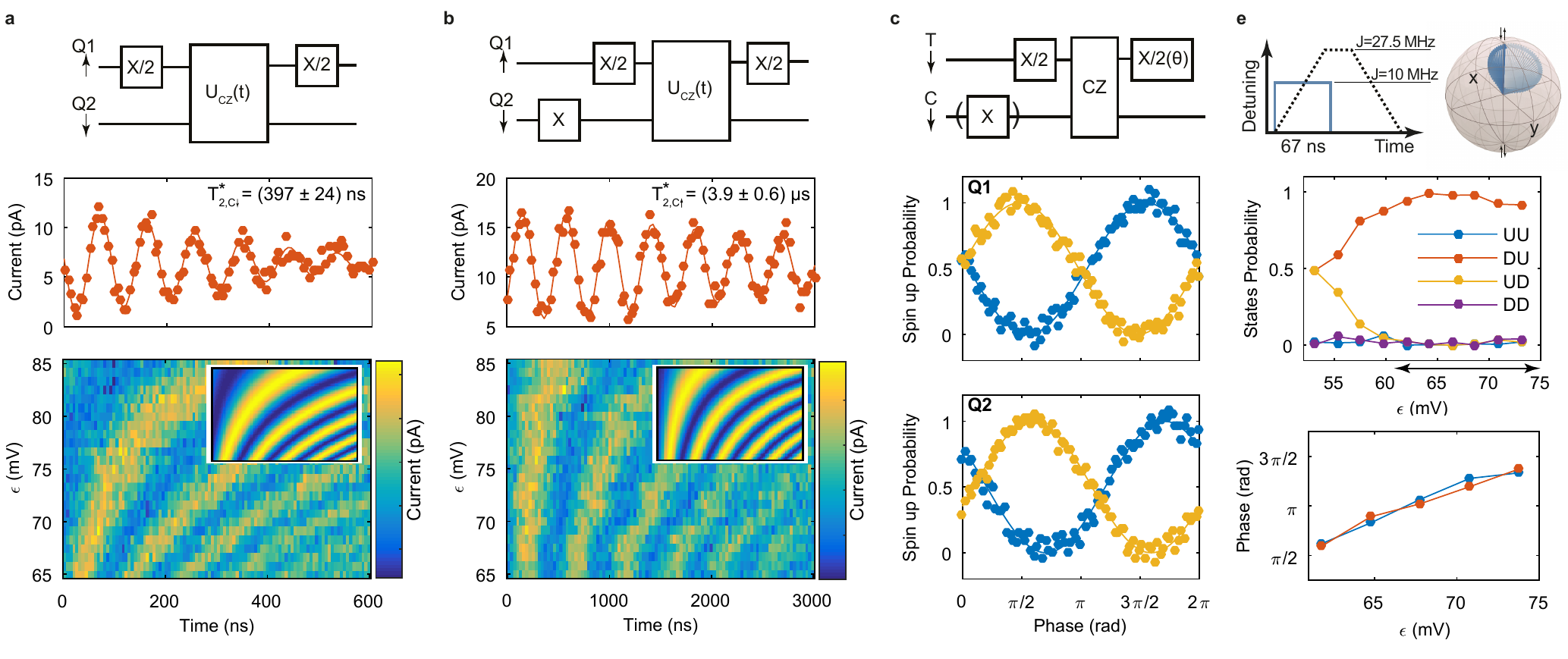}%
	\caption{\textbf{Adiabatic and diabatic CPHASE operation at T = 1.05 K.} 
    	\textbf{a-b} Conditional phase oscillations by adiabatically pulsing the detuning energy $\epsilon$ to increase the exchange interaction $J$, measured using the quantum circuit depicted in the top panels. The antiparallel spin states acquire a phase with respect to the parallel states, resulting in coherent oscillations as a function of the duration of the detuning pulse. At smaller detuning values, the exchange interaction increases resulting in faster oscillations. Due to the exchange interaction, the energy difference $E_{\downarrow\uparrow}-E_{\downarrow\downarrow}$ (measured in \textbf{a}) is smaller than $E_{\uparrow\uparrow}-E_{\uparrow\downarrow}$ (measured in \textbf{b}), resulting in an acquired phase on the target qubit (T) that is dependent on the state of the control qubit (C). 	\textbf{c} Schematic of the quantum circuit to verify CPHASE operation. The adiabatic detuning pulse of the CPHASE gate is tuned such that the antiparallel spin states acquire a total phase of $3\mathrm{\pi}$. The exchange is increased to $J$ = 27.5 MHz using a ramp $t_r$ = 60 ns and the total gate time is $t_\mathrm{CPHASE}$ = 152 ns. We verify CPHASE operation by measuring the normalized spin-up probability, obtained through conversion of the readout current, and observe clear antiparallel oscillations. 
    	\textbf{d} Schematic representation of an adiabatic (dashed black and shown in \textbf{c}) and a diabatic (solid blue) CPHASE. The diabatic CPHASE is optimized by changing the amplitude of $\epsilon$ and measuring probabilities of the four possible spin states. Due to the finite Zeeman difference ($\Delta E_\mathrm{z}=\SI{11}{\mega\hertz}$) SWAP-interactions are not negligible. However, the exchange can be tuned such that the states undergo rotations of $\mathrm{2\pi}$. We tune and optimize this by measuring the phase, projected to the spin states through a $\pi/2$-pulse on the target qubit. We obtain a diabatic CPHASE for $t_\mathrm{CPHASE}$ = 67 ns. 
	}
\label{fig:CPHASE}
\end{figure*}

An alternative way to achieve a universal gate set is through the implementation of the CPHASE gate. Moving in detuning energy toward the (1,5)-(2,4) charge anticrossing lowers the energy of the antiparallel $\ket{\downarrow\uparrow}$ and $\ket{\uparrow\downarrow}$ states with respect to the parallel $\ket{\downarrow\downarrow}$ and $\ket{\uparrow\uparrow}$ spin states. Therefore, pulsing the detuning for a time $t$ results in a phase gate on the target qubit conditional on the spin state of the control qubit. When the total phase $\phi$ = $\phi_{\ket{\downarrow\uparrow}}$ + $\phi_{\ket{\uparrow\downarrow}}$ = $(2n+1)\mathrm{\pi}$ with $n$ integer, a CPHASE gate is realized~\cite{meunier2011efficient}. A high-fidelity implementation of such a gate requires a Zeeman energy difference between the two qubits much larger than the exchange interaction, in order to suppress the evolution of the exchange gate \cite{loss_quantum_1998}. This condition is conveniently met in devices with micromagnets \cite{watson_programmable_2018, zajac_resonantly_2018}, where the CPHASE is the most natural choice as native two-qubit gate.

In our system, $\Delta E_\mathrm{z}$ is comparable in magnitude to the accessible $J$ (see Supplemental Material Sec. II), due to the small $B_\mathrm{ext}$ applied. This means that a detuning pulse will also cause the $\ket{\downarrow\uparrow}$ and $\ket{\uparrow\downarrow}$ states to undergo SWAP rotations. While these rotations occur along a tilted angle due to the non-zero $\Delta E_\mathrm{z}$, they can still reduce the fidelity of the CPHASE gate. In order to avoid unwanted SWAP rotations we implement an adiabatic detuning pulse, by ramping $\epsilon$ to the desired value instead of changing it instantaneously (see schematic in Fig. \ref{fig:CPHASE}e). In this way, a high-fidelity CPHASE gate can still be realized with an arbitrarily small $\Delta E_\mathrm{z}$ at the cost of a longer gate time. In Fig. \ref{fig:CPHASE}a and \ref{fig:CPHASE}b we change the duration of a detuning pulse in between a Ramsey-like experiment on Q1, with and without a $\mathrm{\pi}$ pulse applied to Q2. The frequency of the oscillations of Q1 depends strongly on the spin state of Q2, thereby demonstrating a controlled phase operation. Because of the finite Zeeman energy difference, the antiparallel $\ket{\downarrow\uparrow}$ state shifts significantly more in energy than the $\ket{\uparrow\downarrow}$ state. Consequently, the oscillations in Fig. \ref{fig:CPHASE}a are significantly faster than in Fig. \ref{fig:CPHASE}b. Similarly, the decay time in Fig. \ref{fig:CPHASE}b is significantly longer than in Fig. \ref{fig:CPHASE}a because of the lower sensitivity to electrical noise. In Fig. \ref{fig:CPHASE}c the pulse time is calibrated such that the total phase $\phi$ = $3\mathrm{\pi}$. We measure this in a Ramsey-like experiment where we probe the phase acquired by the target qubit for different control qubit states. From Fig. \ref{fig:CPHASE}c we can observe that the resulting oscillations are nicely out-of-phase, which demonstrates the CPHASE gate. We achieve a gate time $t_\mathrm{CPHASE}$ = \SI{152}{\nano\second}, which is mostly limited by the adiabatic ramps which take $t_\mathrm{r} = \SI{60}{\nano\second}$. From a comparison with simulations we find that the contribution of both ramps to the total phase $\phi$ is approximately $1.7\pi$.

\begin{figure*}[!t]
	\includegraphics[width=\linewidth]{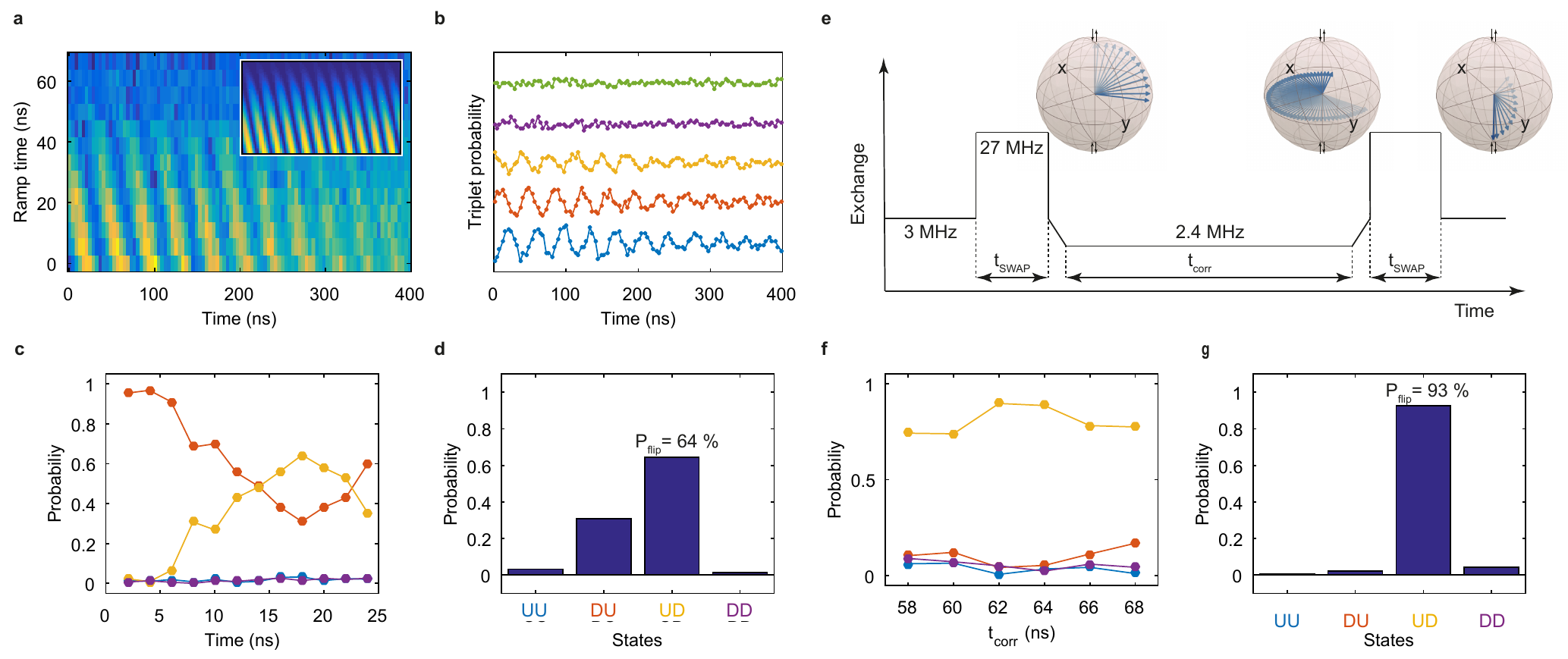}%
	\caption{\textbf{Pulsed SWAP and composite exchange pulse for high-fidelity SWAP at T = 1.05 K.} 
    	\textbf{a-b} SWAP oscillations as a function of the ramp time for a detuning pulse such that $J=\SI{23}{\mega\hertz}$. When the pulsing becomes adiabatic with respect to variations in $J$, the exchange oscillations are suppressed. In order to maximize the readout signal we project the  $\ket{\uparrow\downarrow}$ to the $\ket{\uparrow\uparrow}$ with a $\mathrm{\pi}$ pulse on $\mathrm{f_2}$.   
    	\textbf{c-d} Probabilities of the four spin states as a function of the SWAP interaction time. The states $\ket{\uparrow\uparrow}$ and $\ket{\downarrow\downarrow}$ are not affected, while the states $\ket{\downarrow\uparrow}$ and $\ket{\uparrow\downarrow}$ oscillate. Due to the finite Zeeman difference we achieve a maximum $\ket{\uparrow\downarrow}$ state probability of 64 \% for $t_\mathrm{SWAP}=\SI{18}{\nano\second}$. The exchange interaction is set to $J=\SI{27}{\mega\hertz}$. \textbf{e} Pulse sequence of the composite SWAP gate to correct for errors coming from the finite Zeeman energy difference. The Bloch spheres on top show the time evolution when starting in the $\ket{\downarrow\uparrow}$ state, with the Bloch vector depicted in nanosecond time steps. We first diabatically pulse the exchange to $J=\SI{27}{\mega\hertz}$, in order to bring the state on the equator of the singlet-triplet Bloch sphere. Then we correct for the phase offset with an adiabatic exchange pulse to $J=\SI{2.4}{\mega\hertz}$. We complete the state flip with another exchange pulse to $J=\SI{27}{\mega\hertz}$.
    	\textbf{f} Spin state probability after applying the composite SWAP and as a function of the adiabatic pulse time $t_\mathrm{corr}$, from which we find the optimum $t_\mathrm{corr}=\SI{62}{\nano\second}$.
    	\textbf{g} Spin state probability after executing the composite SWAP sequence starting from the initial state $\ket{\downarrow\uparrow}$. Compared to the detuning pulse as shown in \textbf{d} we find a clear improvement in the spin flip SWAP probability. 
	}
\label{fig:SWAP}
\end{figure*}

This gate time can be significantly sped up with the implementation of a geometric CPHASE gate, that does not require adiabaticity~\cite{burkard1999physical}. For the implementation of this gate we synchronize the unwanted exchange oscillations with the total gate duration, i.e. our gate performs a CPHASE evolution while the exchange oscillations performs a complete cycle. For a perfectly diabatic pulse the condition for the exchange interaction is: 
\begin{align}
J=(4\,J_\mathrm{res}+\sqrt{3\Delta E_\mathrm{z}^2+4J_\mathrm{res}^2})/3,
\label{eq:geom_cphase}
\end{align}
where $J_\mathrm{res}$ is the residual exchange interaction at the point where we perform CROT gates (see Supplemental Material Sec. III). 

Figures \ref{fig:CPHASE}e show the experimental implementation of the geometric CPHASE gate. We sweep the amplitude of the detuning pulse and monitor the spin state probabilities (see Supplemental Material Sec. I) during exchange oscillations, and the total phase acquired by the antiparallel spin states. We notice that, when $\epsilon \approx \SI{68}{\milli\volt}$, the antiparallel spin states execute a $\mathrm{2\pi}$ rotation, while acquiring a total phase shift of $\pi$. At this value of detuning we measure $J\approx \SI{10}{MHz}$ (see Supplemental Material Sec. II) and therefore in agreement with Eq.\ref{eq:geom_cphase}. The total gate time is reduced here to $t_\mathrm{CPHASE}=\SI{67}{\nano\second}$.

We now turn to the implementation of a SWAP gate, the originally proposed quantum gate for quantum dots~\cite{loss_quantum_1998}. Despite the experimental demonstration of exchange oscillations \cite{petta_coherent_2005, maune2012coherent, he2019two}, its implementation together with single-qubit gates is rather challenging because of the requirement of a negligible Zeeman difference between the qubits. In the following we will discuss a novel protocol that can overcome this problem and allow for a high-fidelity SWAP gate, even in the presence of a finite $\Delta E_\mathrm{z}$.

In order to observe SWAP oscillations, we implement a sequence where we initialize in the $\ket{\downarrow\uparrow}$ state and pulse $\epsilon$ for a time $t$. Clear exchange oscillations between the $\ket{\downarrow\uparrow}$ and the $\ket{\uparrow\downarrow}$ state are visible when the detuning pulse is diabatic (see Fig. \ref{fig:SWAP}a and \ref{fig:SWAP}b), where the oscillation frequency is $f_\mathrm{SWAP} = \sqrt{J^2 + \Delta E_\mathrm{z}^2}$. As we make the pulse more adiabatic by ramping $\epsilon$, the oscillations disappear and the regime becomes suitable for a CPHASE implementation as discussed before. Even when the detuning pulse is perfectly diabatic, we do not obtain a perfect SWAP due to the finite $\Delta E_\mathrm{z}$. Instead, the spin states rotate in the Bloch sphere around the tilted axis of rotation $\boldsymbol{r}=(J,0,\Delta E_\text{z})^T$, similar to what happens for off-resonant driving. Figure \ref{fig:SWAP}c and \ref{fig:SWAP}d show that when starting in the $\ket{\downarrow\uparrow}$ state, a maximum $\ket{\uparrow\downarrow}$ state probability of $64 \%$ is obtained in $t_\mathrm{SWAP}=\SI{18}{\nano\second}$, which is in agreement with our simulated predictions (see Supplemental Material Sec. III). 

\begin{figure}[!t]
	\includegraphics[width=\linewidth]{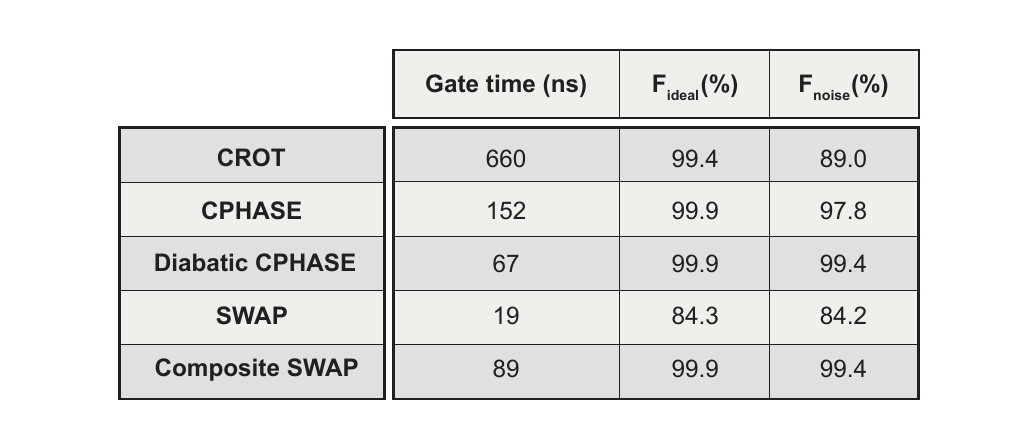}%
	\caption{\textbf{Gate times and simulated fidelities for silicon qubits at T = 1.05 K.} 
    	Gate times and simulated fidelities for all the two-qubit gates discussed in the main text, where $F_{ideal}$ represent the fidelity in the absence of noise and $F_{noise}$ takes into account the experimental noise at 1.05 Kelvin. We find high-fidelity two-qubit gates can be obtained in silicon above one Kelvin, by using diabatic CPHASE or composite SWAP sequences. The CROT fidelity is calculated as a conditional $\pi$-flip for better comparison. Good agreement is obtained with previous experiments~\cite{petit2020universal}, confirming that the simulated noise is an accurate estimate of the real noise. Further improvement in the fidelities of the CROT and the CPHASE may be obtained by incorporating pulse shaping \cite{martinis2014fast,gungordu2018pulse, calderon2019fast, gungordu2019analytically, gungordu2020robust}. 
	}
\label{fig:table}
\end{figure}

Composite pulse sequences~\cite{vandersypen2005nmr} can correct for the tilted axis of rotation. It is possible to achieve full population transfer with an exchange sequence consisting of alternating diabatic and adiabatic exchange pulses. The corresponding time evolution operators in the odd parity subspace are:

\begin{align}
U_\text{r}=e^{i \Phi_\text{r}}e^{i \theta_\text{r} \boldsymbol{r}\cdot\boldsymbol{\sigma}} \\
U_\text{z}=e^{i \Phi_\text{z}}e^{i \theta_\text{z} \hat{Z}} 
\end{align}

for a diabatic and an adiabatic pulse respectively (see Supplemental Material Sec. III). Here $\boldsymbol{\sigma}=(\hat{X},\hat{Y},\hat{Z})$ is the vector consisting of the Pauli matrices, $\Phi_{\text{r},\text{z}}=J t_{\text{r},\text{z}}/2$ the accumulated entangling phase during the pulse, and $\theta_{\text{r},\text{z}} = t_{\text{r},\text{z}} \sqrt{J^2+\Delta E_\text{z}^2}/2$ the angle of rotation. The condition for a SWAP gate is $U_\text{tot}= U_\text{r}U_\text{z}U_\text{r}U_{\text{z}_2}U_{\text{r}_2} \cdots\equiv \hat{X}$. The number of necessary pulses depends on the angle of rotation; obviously a minimal pulse sequence requires $|\Delta E_\text{z}| \leq  J$. Furthermore, it is essential to include the global phase which corresponds to a conditional phase evolution in the full two-qubit space and needs to vanish when implementing a SWAP gate. This protocol is highly versatile and can also produce maximally entangling gates, i.e., $\sqrt{\text{SWAP}}$ if $U_\text{tot}\equiv i\hat{X}/2$ and $i\text{SWAP}$ for $U_\text{tot}\equiv i\hat{X}$.

A possible minimal length solution for a SWAP gate is sketched in Fig. \ref{fig:SWAP}e and the trajectory of the qubit state is seen in the inset. In the experiment, we calibrate the exchange interaction at all stages of the pulse, fix the time of the diabatic pulses to 12 ns and sweep the length of the adiabatic pulse $t_\mathrm{corr}$ in order to find the best point. Figure \ref{fig:SWAP}f shows how the four spin probabilities change when sweeping $t_\mathrm{corr}$. We find an optimal $t_\mathrm{corr}=\SI{62}{\nano\second}$ and the four spin state probabilities for a total pulse duration $t_\mathrm{SWAP}=\SI{88}{\nano\second}$ are plotted in Fig. \ref{fig:SWAP}g. The SWAP probability exceeds $90\%$, where the remaining error is dominated by miscalibrations, inaccuracies in the gates needed to reconstruct the spin state probabilities, and state-preparation-and-measurement (SPAM) errors.

Table \ref{fig:table} shows the fidelities associated with the two-qubit gates CROT, CPHASE, and SWAP. Here, $F_\mathrm{ideal}$ represents the simulated fidelities taking into account the relevant parameters, but neglecting any decoherence. We find  $F_\mathrm{ideal}>99\%$ for all gates except the SWAP, which is limited in fidelity  by the finite $\Delta E_\mathrm{z}$. We have also modelled the decoherence assuming $1/f$ noise as the main noise source (see Supplemental Material Sec. III). By fitting the experimental data in Fig \ref{fig:CPHASE}a and \ref{fig:CPHASE}b, we conclude that our model is able to reproduce the decoherence with good agreement. Based on these simulations we determine $F_\mathrm{noise}$. The fidelity of the CROT and the CPHASE gate are significantly affected by the noise, due to the relatively long gate times, and we find that the predicted CROT fidelity $F_\mathrm{noise}$ = 89 $\%$ is close to the experimentally measured fidelity $F$ = 86 $\%$ \cite{petit2020universal}. The SWAP, diabatic CPHASE and composite SWAP are less affected by the noise and in particular we predict that both the diabatic CPHASE and composite SWAP can be executed with fidelities above 99 $\%$. 

The ability to execute a diverse set of high-fidelity two-qubit gates define silicon quantum dots as a versatile platform for quantum information. The low magnetic field operation and the small Zeeman energy difference between qubits is furthermore beneficial for the realization of scalable qubit tiles, as it supports high-fidelity shuttlers and on-chip resonators for long-distance qubit links. Moreover, the ability to execute quantum logic at temperatures exceeding one Kelvin provides a pathway to quantum integrated circuits that host both the qubits and their control circuitry for scalable quantum hardware. 

\section*{Data availability}
All data underlying this study will be available from the 4TU ResearchData repository.

\section*{Acknowledgements}
We thank S. de Snoo and S.G.J. Philips for software developments and the Veldhorst group for useful discussions. M. V. acknowledges support through a Vidi and an NWO grant, both associated with the Netherlands Organization of Scientific Research (NWO). Research was sponsored by the Army Research Office (ARO) and was accomplished under Grant No. W911NF- 17-1-0274. The views and conclusions contained in this document are those of the authors and should not be interpreted as representing the official policies, either expressed or implied, of the Army Research Office (ARO), or the U.S. Government. The U.S. Government is authorized to reproduce and distribute reprints for Government purposes notwithstanding any copyright notation herein. 

\section*{Authors contributions}
L.P. performed the experiments, M.R developed the theoretical models. H.G.J.E. and W.I.L.L. were responsible for the device fabrication. J.S.C. supervised the wafer growth. L.P. and M.R. analysed the data with input from L.M.K.V. and M.V. L.P., M.R. and M.V. wrote the manuscript with input from all authors. M.V. and L.M.K.V acquired the funding. M.V. conceived and supervised the project.

\section*{Competing Interests}
The authors declare no competing interests. Correspondence should be addressed to M.V. (M.Veldhorst@tudelft.nl).


\begin{thebibliography}{34}%
	\makeatletter
	\providecommand \@ifxundefined [1]{%
		\@ifx{#1\undefined}
	}%
	\providecommand \@ifnum [1]{%
		\ifnum #1\expandafter \@firstoftwo
		\else \expandafter \@secondoftwo
		\fi
	}%
	\providecommand \@ifx [1]{%
		\ifx #1\expandafter \@firstoftwo
		\else \expandafter \@secondoftwo
		\fi
	}%
	\providecommand \natexlab [1]{#1}%
	\providecommand \enquote  [1]{``#1''}%
	\providecommand \bibnamefont  [1]{#1}%
	\providecommand \bibfnamefont [1]{#1}%
	\providecommand \citenamefont [1]{#1}%
	\providecommand \href@noop [0]{\@secondoftwo}%
	\providecommand \href [0]{\begingroup \@sanitize@url \@href}%
	\providecommand \@href[1]{\@@startlink{#1}\@@href}%
	\providecommand \@@href[1]{\endgroup#1\@@endlink}%
	\providecommand \@sanitize@url [0]{\catcode `\\12\catcode `\$12\catcode
		`\&12\catcode `\#12\catcode `\^12\catcode `\_12\catcode `\%12\relax}%
	\providecommand \@@startlink[1]{}%
	\providecommand \@@endlink[0]{}%
	\providecommand \url  [0]{\begingroup\@sanitize@url \@url }%
	\providecommand \@url [1]{\endgroup\@href {#1}{\urlprefix }}%
	\providecommand \urlprefix  [0]{URL }%
	\providecommand \Eprint [0]{\href }%
	\providecommand \doibase [0]{http://dx.doi.org/}%
	\providecommand \selectlanguage [0]{\@gobble}%
	\providecommand \bibinfo  [0]{\@secondoftwo}%
	\providecommand \bibfield  [0]{\@secondoftwo}%
	\providecommand \translation [1]{[#1]}%
	\providecommand \BibitemOpen [0]{}%
	\providecommand \bibitemStop [0]{}%
	\providecommand \bibitemNoStop [0]{.\EOS\space}%
	\providecommand \EOS [0]{\spacefactor3000\relax}%
	\providecommand \BibitemShut  [1]{\csname bibitem#1\endcsname}%
	\let\auto@bib@innerbib\@empty
	\bibitem [{\citenamefont {Loss}\ and\ \citenamefont
		{DiVincenzo}(1998)}]{loss_quantum_1998}%
	\BibitemOpen
	\bibfield  {author} {\bibinfo {author} {\bibfnamefont {D.}~\bibnamefont
			{Loss}}\ and\ \bibinfo {author} {\bibfnamefont {D.~P.}\ \bibnamefont
			{DiVincenzo}},\ }\href {\doibase 10.1103/PhysRevA.57.120} {\bibfield
		{journal} {\bibinfo  {journal} {Physical Review A}\ }\textbf {\bibinfo
			{volume} {57}},\ \bibinfo {pages} {120} (\bibinfo {year} {1998})}\BibitemShut
	{NoStop}%
	\bibitem [{\citenamefont {Vandersypen}\ \emph {et~al.}(2017)\citenamefont
		{Vandersypen}, \citenamefont {Bluhm}, \citenamefont {Clarke}, \citenamefont
		{Dzurak}, \citenamefont {Ishihara}, \citenamefont {Morello}, \citenamefont
		{Reilly}, \citenamefont {Schreiber},\ and\ \citenamefont
		{Veldhorst}}]{vandersypen_interfacing_2017}%
	\BibitemOpen
	\bibfield  {author} {\bibinfo {author} {\bibfnamefont {L.~M.~K.}\
			\bibnamefont {Vandersypen}}, \bibinfo {author} {\bibfnamefont
			{H.}~\bibnamefont {Bluhm}}, \bibinfo {author} {\bibfnamefont {J.~S.}\
			\bibnamefont {Clarke}}, \bibinfo {author} {\bibfnamefont {A.~S.}\
			\bibnamefont {Dzurak}}, \bibinfo {author} {\bibfnamefont {R.}~\bibnamefont
			{Ishihara}}, \bibinfo {author} {\bibfnamefont {A.}~\bibnamefont {Morello}},
		\bibinfo {author} {\bibfnamefont {D.~J.}\ \bibnamefont {Reilly}}, \bibinfo
		{author} {\bibfnamefont {L.~R.}\ \bibnamefont {Schreiber}}, \ and\ \bibinfo
		{author} {\bibfnamefont {M.}~\bibnamefont {Veldhorst}},\ }\href {\doibase
		10.1038/s41534-017-0038-y} {\bibfield  {journal} {\bibinfo  {journal} {npj
				Quantum Information}\ }\textbf {\bibinfo {volume} {3}},\ \bibinfo {pages}
		{34} (\bibinfo {year} {2017})}\BibitemShut {NoStop}%
	\bibitem [{\citenamefont {Veldhorst}\ \emph {et~al.}(2014)\citenamefont
		{Veldhorst}, \citenamefont {Hwang}, \citenamefont {Yang}, \citenamefont
		{Leenstra}, \citenamefont {de~Ronde}, \citenamefont {Dehollain},
		\citenamefont {Muhonen}, \citenamefont {Hudson}, \citenamefont {Itoh},
		\citenamefont {Morello},\ and\ \citenamefont
		{Dzurak}}]{veldhorst_addressable_2014}%
	\BibitemOpen
	\bibfield  {author} {\bibinfo {author} {\bibfnamefont {M.}~\bibnamefont
			{Veldhorst}}, \bibinfo {author} {\bibfnamefont {J.~C.~C.}\ \bibnamefont
			{Hwang}}, \bibinfo {author} {\bibfnamefont {C.~H.}\ \bibnamefont {Yang}},
		\bibinfo {author} {\bibfnamefont {A.~W.}\ \bibnamefont {Leenstra}}, \bibinfo
		{author} {\bibfnamefont {B.}~\bibnamefont {de~Ronde}}, \bibinfo {author}
		{\bibfnamefont {J.~P.}\ \bibnamefont {Dehollain}}, \bibinfo {author}
		{\bibfnamefont {J.~T.}\ \bibnamefont {Muhonen}}, \bibinfo {author}
		{\bibfnamefont {F.~E.}\ \bibnamefont {Hudson}}, \bibinfo {author}
		{\bibfnamefont {K.~M.}\ \bibnamefont {Itoh}}, \bibinfo {author}
		{\bibfnamefont {A.}~\bibnamefont {Morello}}, \ and\ \bibinfo {author}
		{\bibfnamefont {A.~S.}\ \bibnamefont {Dzurak}},\ }\href@noop {} {\bibfield
		{journal} {\bibinfo  {journal} {Nature Nanotechnology}\ }\textbf {\bibinfo
			{volume} {9}},\ \bibinfo {pages} {981} (\bibinfo {year} {2014})}\BibitemShut
	{NoStop}%
	\bibitem [{\citenamefont {Yang}\ \emph {et~al.}(2020)\citenamefont {Yang},
		\citenamefont {Leon}, \citenamefont {Hwang}, \citenamefont {Saraiva},
		\citenamefont {Tanttu}, \citenamefont {Huang}, \citenamefont {Lemyre},
		\citenamefont {Chan}, \citenamefont {Tan}, \citenamefont {Hudson},
		\citenamefont {Itoh}, \citenamefont {Morello}, \citenamefont
		{Pioro-Ladriere}, \citenamefont {Laucht},\ and\ \citenamefont
		{Dzurak}}]{yang2020operation}%
	\BibitemOpen
	\bibfield  {author} {\bibinfo {author} {\bibfnamefont {C.~H.}\ \bibnamefont
			{Yang}}, \bibinfo {author} {\bibfnamefont {R.~C.~C.}\ \bibnamefont {Leon}},
		\bibinfo {author} {\bibfnamefont {J.~C.~C.}\ \bibnamefont {Hwang}}, \bibinfo
		{author} {\bibfnamefont {A.}~\bibnamefont {Saraiva}}, \bibinfo {author}
		{\bibfnamefont {T.}~\bibnamefont {Tanttu}}, \bibinfo {author} {\bibfnamefont
			{W.}~\bibnamefont {Huang}}, \bibinfo {author} {\bibfnamefont {J.~C.}\
			\bibnamefont {Lemyre}}, \bibinfo {author} {\bibfnamefont {K.~W.}\
			\bibnamefont {Chan}}, \bibinfo {author} {\bibfnamefont {K.~Y.}\ \bibnamefont
			{Tan}}, \bibinfo {author} {\bibfnamefont {F.~E.}\ \bibnamefont {Hudson}},
		\bibinfo {author} {\bibfnamefont {K.~M.}\ \bibnamefont {Itoh}}, \bibinfo
		{author} {\bibfnamefont {A.}~\bibnamefont {Morello}}, \bibinfo {author}
		{\bibfnamefont {M.}~\bibnamefont {Pioro-Ladriere}}, \bibinfo {author}
		{\bibfnamefont {A.}~\bibnamefont {Laucht}}, \ and\ \bibinfo {author}
		{\bibfnamefont {A.~S.}\ \bibnamefont {Dzurak}},\ }\href@noop {} {\bibfield
		{journal} {\bibinfo  {journal} {Nature}\ }\textbf {\bibinfo {volume} {580}},\
		\bibinfo {pages} {350} (\bibinfo {year} {2020})}\BibitemShut {NoStop}%
	\bibitem [{\citenamefont {Petit}\ \emph {et~al.}(2020)\citenamefont {Petit},
		\citenamefont {Eenink}, \citenamefont {Russ}, \citenamefont {Lawrie},
		\citenamefont {Hendrickx}, \citenamefont {Philips}, \citenamefont {Clarke},
		\citenamefont {Vandersypen},\ and\ \citenamefont
		{Veldhorst}}]{petit2020universal}%
	\BibitemOpen
	\bibfield  {author} {\bibinfo {author} {\bibfnamefont {L.}~\bibnamefont
			{Petit}}, \bibinfo {author} {\bibfnamefont {H.~G.~J.}\ \bibnamefont
			{Eenink}}, \bibinfo {author} {\bibfnamefont {M.}~\bibnamefont {Russ}},
		\bibinfo {author} {\bibfnamefont {W.~I.~L.}\ \bibnamefont {Lawrie}}, \bibinfo
		{author} {\bibfnamefont {N.~W.}\ \bibnamefont {Hendrickx}}, \bibinfo {author}
		{\bibfnamefont {S.~G.~J.}\ \bibnamefont {Philips}}, \bibinfo {author}
		{\bibfnamefont {J.~S.}\ \bibnamefont {Clarke}}, \bibinfo {author}
		{\bibfnamefont {L.~M.~K.}\ \bibnamefont {Vandersypen}}, \ and\ \bibinfo
		{author} {\bibfnamefont {M.}~\bibnamefont {Veldhorst}},\ }\href@noop {}
	{\bibfield  {journal} {\bibinfo  {journal} {Nature}\ }\textbf {\bibinfo
			{volume} {580}},\ \bibinfo {pages} {355} (\bibinfo {year}
		{2020})}\BibitemShut {NoStop}%
	\bibitem [{\citenamefont {Petta}\ \emph {et~al.}(2005)\citenamefont {Petta},
		\citenamefont {Johnson}, \citenamefont {Taylor}, \citenamefont {Laird},
		\citenamefont {Yacoby}, \citenamefont {Lukin}, \citenamefont {Marcus},
		\citenamefont {Hanson},\ and\ \citenamefont {Gossard}}]{petta_coherent_2005}%
	\BibitemOpen
	\bibfield  {author} {\bibinfo {author} {\bibfnamefont {J.~R.}\ \bibnamefont
			{Petta}}, \bibinfo {author} {\bibfnamefont {A.~C.}\ \bibnamefont {Johnson}},
		\bibinfo {author} {\bibfnamefont {J.~M.}\ \bibnamefont {Taylor}}, \bibinfo
		{author} {\bibfnamefont {E.~A.}\ \bibnamefont {Laird}}, \bibinfo {author}
		{\bibfnamefont {A.}~\bibnamefont {Yacoby}}, \bibinfo {author} {\bibfnamefont
			{M.~D.}\ \bibnamefont {Lukin}}, \bibinfo {author} {\bibfnamefont {C.~M.}\
			\bibnamefont {Marcus}}, \bibinfo {author} {\bibfnamefont {M.~P.}\
			\bibnamefont {Hanson}}, \ and\ \bibinfo {author} {\bibfnamefont {A.~C.}\
			\bibnamefont {Gossard}},\ }\href {\doibase 10.1126/science.1116955}
	{\bibfield  {journal} {\bibinfo  {journal} {Science}\ }\textbf {\bibinfo
			{volume} {309}},\ \bibinfo {pages} {2180} (\bibinfo {year}
		{2005})}\BibitemShut {NoStop}%
	\bibitem [{\citenamefont {Veldhorst}\ \emph {et~al.}(2015)\citenamefont
		{Veldhorst}, \citenamefont {Yang}, \citenamefont {Hwang}, \citenamefont
		{Huang}, \citenamefont {Dehollain}, \citenamefont {Muhonen}, \citenamefont
		{Simmons}, \citenamefont {Laucht}, \citenamefont {Hudson}, \citenamefont
		{Itoh}, \citenamefont {Morello},\ and\ \citenamefont
		{Dzurak}}]{veldhorst_two-qubit_2015}%
	\BibitemOpen
	\bibfield  {author} {\bibinfo {author} {\bibfnamefont {M.}~\bibnamefont
			{Veldhorst}}, \bibinfo {author} {\bibfnamefont {C.~H.}\ \bibnamefont {Yang}},
		\bibinfo {author} {\bibfnamefont {J.~C.~C.}\ \bibnamefont {Hwang}}, \bibinfo
		{author} {\bibfnamefont {W.}~\bibnamefont {Huang}}, \bibinfo {author}
		{\bibfnamefont {J.~P.}\ \bibnamefont {Dehollain}}, \bibinfo {author}
		{\bibfnamefont {J.~T.}\ \bibnamefont {Muhonen}}, \bibinfo {author}
		{\bibfnamefont {S.}~\bibnamefont {Simmons}}, \bibinfo {author} {\bibfnamefont
			{A.}~\bibnamefont {Laucht}}, \bibinfo {author} {\bibfnamefont {F.~E.}\
			\bibnamefont {Hudson}}, \bibinfo {author} {\bibfnamefont {K.~M.}\
			\bibnamefont {Itoh}}, \bibinfo {author} {\bibfnamefont {A.}~\bibnamefont
			{Morello}}, \ and\ \bibinfo {author} {\bibfnamefont {A.~S.}\ \bibnamefont
			{Dzurak}},\ }\href {\doibase 10.1038/nature15263} {\bibfield  {journal}
		{\bibinfo  {journal} {Nature}\ }\textbf {\bibinfo {volume} {526}},\ \bibinfo
		{pages} {410} (\bibinfo {year} {2015})}\BibitemShut {NoStop}%
	\bibitem [{\citenamefont {Watson}\ \emph {et~al.}(2018)\citenamefont {Watson},
		\citenamefont {Philips}, \citenamefont {Kawakami}, \citenamefont {Ward},
		\citenamefont {Scarlino}, \citenamefont {Veldhorst}, \citenamefont {Savage},
		\citenamefont {Lagally}, \citenamefont {Friesen}, \citenamefont
		{Coppersmith}, \citenamefont {Eriksson},\ and\ \citenamefont
		{Vandersypen}}]{watson_programmable_2018}%
	\BibitemOpen
	\bibfield  {author} {\bibinfo {author} {\bibfnamefont {T.~F.}\ \bibnamefont
			{Watson}}, \bibinfo {author} {\bibfnamefont {S.~G.~J.}\ \bibnamefont
			{Philips}}, \bibinfo {author} {\bibfnamefont {E.}~\bibnamefont {Kawakami}},
		\bibinfo {author} {\bibfnamefont {D.~R.}\ \bibnamefont {Ward}}, \bibinfo
		{author} {\bibfnamefont {P.}~\bibnamefont {Scarlino}}, \bibinfo {author}
		{\bibfnamefont {M.}~\bibnamefont {Veldhorst}}, \bibinfo {author}
		{\bibfnamefont {D.~E.}\ \bibnamefont {Savage}}, \bibinfo {author}
		{\bibfnamefont {M.~G.}\ \bibnamefont {Lagally}}, \bibinfo {author}
		{\bibfnamefont {M.}~\bibnamefont {Friesen}}, \bibinfo {author} {\bibfnamefont
			{S.~N.}\ \bibnamefont {Coppersmith}}, \bibinfo {author} {\bibfnamefont
			{M.~A.}\ \bibnamefont {Eriksson}}, \ and\ \bibinfo {author} {\bibfnamefont
			{L.~M.~K.}\ \bibnamefont {Vandersypen}},\ }\href {\doibase
		10.1038/nature25766} {\bibfield  {journal} {\bibinfo  {journal} {Nature}\
		}\textbf {\bibinfo {volume} {555}},\ \bibinfo {pages} {633} (\bibinfo {year}
		{2018})}\BibitemShut {NoStop}%
	\bibitem [{\citenamefont {Koppens}\ \emph {et~al.}(2006)\citenamefont
		{Koppens}, \citenamefont {Buizert}, \citenamefont {Tielrooij}, \citenamefont
		{Vink}, \citenamefont {Nowack}, \citenamefont {Meunier}, \citenamefont
		{Kouwenhoven},\ and\ \citenamefont {Vandersypen}}]{koppens2006driven}%
	\BibitemOpen
	\bibfield  {author} {\bibinfo {author} {\bibfnamefont {F.~H.~L.}\
			\bibnamefont {Koppens}}, \bibinfo {author} {\bibfnamefont {C.}~\bibnamefont
			{Buizert}}, \bibinfo {author} {\bibfnamefont {K.-J.}\ \bibnamefont
			{Tielrooij}}, \bibinfo {author} {\bibfnamefont {I.~T.}\ \bibnamefont {Vink}},
		\bibinfo {author} {\bibfnamefont {K.~C.}\ \bibnamefont {Nowack}}, \bibinfo
		{author} {\bibfnamefont {T.}~\bibnamefont {Meunier}}, \bibinfo {author}
		{\bibfnamefont {L.~P.}\ \bibnamefont {Kouwenhoven}}, \ and\ \bibinfo {author}
		{\bibfnamefont {L.~M.~K.}\ \bibnamefont {Vandersypen}},\ }\href@noop {}
	{\bibfield  {journal} {\bibinfo  {journal} {Nature}\ }\textbf {\bibinfo
			{volume} {442}},\ \bibinfo {pages} {766} (\bibinfo {year}
		{2006})}\BibitemShut {NoStop}%
	\bibitem [{\citenamefont {Zajac}\ \emph {et~al.}(2018)\citenamefont {Zajac},
		\citenamefont {Sigillito}, \citenamefont {Russ}, \citenamefont {Borjans},
		\citenamefont {Taylor}, \citenamefont {Burkard},\ and\ \citenamefont
		{Petta}}]{zajac_resonantly_2018}%
	\BibitemOpen
	\bibfield  {author} {\bibinfo {author} {\bibfnamefont {D.~M.}\ \bibnamefont
			{Zajac}}, \bibinfo {author} {\bibfnamefont {A.~J.}\ \bibnamefont
			{Sigillito}}, \bibinfo {author} {\bibfnamefont {M.}~\bibnamefont {Russ}},
		\bibinfo {author} {\bibfnamefont {F.}~\bibnamefont {Borjans}}, \bibinfo
		{author} {\bibfnamefont {J.~M.}\ \bibnamefont {Taylor}}, \bibinfo {author}
		{\bibfnamefont {G.}~\bibnamefont {Burkard}}, \ and\ \bibinfo {author}
		{\bibfnamefont {J.~R.}\ \bibnamefont {Petta}},\ }\href {\doibase
		10.1126/science.aao5965} {\bibfield  {journal} {\bibinfo  {journal}
			{Science}\ }\textbf {\bibinfo {volume} {359}},\ \bibinfo {pages} {439}
		(\bibinfo {year} {2018})}\BibitemShut {NoStop}%
	\bibitem [{\citenamefont {Huang}\ \emph {et~al.}(2019)\citenamefont {Huang},
		\citenamefont {Yang}, \citenamefont {Chan}, \citenamefont {Tanttu},
		\citenamefont {Hensen}, \citenamefont {Leon}, \citenamefont {Fogarty},
		\citenamefont {Hwang}, \citenamefont {Hudson}, \citenamefont {Itoh},
		\citenamefont {Morello}, \citenamefont {Laucht},\ and\ \citenamefont
		{Dzurak}}]{huang2019fidelity}%
	\BibitemOpen
	\bibfield  {author} {\bibinfo {author} {\bibfnamefont {W.}~\bibnamefont
			{Huang}}, \bibinfo {author} {\bibfnamefont {C.~H.}\ \bibnamefont {Yang}},
		\bibinfo {author} {\bibfnamefont {K.~W.}\ \bibnamefont {Chan}}, \bibinfo
		{author} {\bibfnamefont {T.}~\bibnamefont {Tanttu}}, \bibinfo {author}
		{\bibfnamefont {B.}~\bibnamefont {Hensen}}, \bibinfo {author} {\bibfnamefont
			{R.~C.~C.}\ \bibnamefont {Leon}}, \bibinfo {author} {\bibfnamefont {M.~A.}\
			\bibnamefont {Fogarty}}, \bibinfo {author} {\bibfnamefont {J.~C.~C.}\
			\bibnamefont {Hwang}}, \bibinfo {author} {\bibfnamefont {F.~E.}\ \bibnamefont
			{Hudson}}, \bibinfo {author} {\bibfnamefont {K.~M.}\ \bibnamefont {Itoh}},
		\bibinfo {author} {\bibfnamefont {A.}~\bibnamefont {Morello}}, \bibinfo
		{author} {\bibfnamefont {A.}~\bibnamefont {Laucht}}, \ and\ \bibinfo {author}
		{\bibfnamefont {A.~S.}\ \bibnamefont {Dzurak}},\ }\href@noop {} {\bibfield
		{journal} {\bibinfo  {journal} {Nature}\ }\textbf {\bibinfo {volume} {569}},\
		\bibinfo {pages} {532} (\bibinfo {year} {2019})}\BibitemShut {NoStop}%
	\bibitem [{\citenamefont {Hendrickx}\ \emph {et~al.}(2020)\citenamefont
		{Hendrickx}, \citenamefont {Franke}, \citenamefont {Sammak}, \citenamefont
		{Scappucci},\ and\ \citenamefont {Veldhorst}}]{hendrickx2020fast}%
	\BibitemOpen
	\bibfield  {author} {\bibinfo {author} {\bibfnamefont {N.~W.}\ \bibnamefont
			{Hendrickx}}, \bibinfo {author} {\bibfnamefont {D.~P.}\ \bibnamefont
			{Franke}}, \bibinfo {author} {\bibfnamefont {A.}~\bibnamefont {Sammak}},
		\bibinfo {author} {\bibfnamefont {G.}~\bibnamefont {Scappucci}}, \ and\
		\bibinfo {author} {\bibfnamefont {M.}~\bibnamefont {Veldhorst}},\ }\href@noop
	{} {\bibfield  {journal} {\bibinfo  {journal} {Nature}\ }\textbf {\bibinfo
			{volume} {577}},\ \bibinfo {pages} {487} (\bibinfo {year}
		{2020})}\BibitemShut {NoStop}%
	\bibitem [{\citenamefont {Arute}\ \emph {et~al.}(2019)\citenamefont {Arute},
		\citenamefont {Arya}, \citenamefont {Babbush}, \citenamefont {Bacon},
		\citenamefont {Bardin}, \citenamefont {Barends}, \citenamefont {Biswas},
		\citenamefont {Boixo}, \citenamefont {Brandao}, \citenamefont {Buell} \emph
		{et~al.}}]{arute2019quantum}%
	\BibitemOpen
	\bibfield  {author} {\bibinfo {author} {\bibfnamefont {F.}~\bibnamefont
			{Arute}}, \bibinfo {author} {\bibfnamefont {K.}~\bibnamefont {Arya}},
		\bibinfo {author} {\bibfnamefont {R.}~\bibnamefont {Babbush}}, \bibinfo
		{author} {\bibfnamefont {D.}~\bibnamefont {Bacon}}, \bibinfo {author}
		{\bibfnamefont {J.~C.}\ \bibnamefont {Bardin}}, \bibinfo {author}
		{\bibfnamefont {R.}~\bibnamefont {Barends}}, \bibinfo {author} {\bibfnamefont
			{R.}~\bibnamefont {Biswas}}, \bibinfo {author} {\bibfnamefont
			{S.}~\bibnamefont {Boixo}}, \bibinfo {author} {\bibfnamefont {F.~G.}\
			\bibnamefont {Brandao}}, \bibinfo {author} {\bibfnamefont {D.~A.}\
			\bibnamefont {Buell}},  \emph {et~al.},\ }\href@noop {} {\bibfield  {journal}
		{\bibinfo  {journal} {Nature}\ }\textbf {\bibinfo {volume} {574}},\ \bibinfo
		{pages} {505} (\bibinfo {year} {2019})}\BibitemShut {NoStop}%
	\bibitem [{\citenamefont {Reiher}\ \emph {et~al.}(2017)\citenamefont {Reiher},
		\citenamefont {Wiebe}, \citenamefont {Svore}, \citenamefont {Wecker},\ and\
		\citenamefont {Troyer}}]{reiher2017elucidating}%
	\BibitemOpen
	\bibfield  {author} {\bibinfo {author} {\bibfnamefont {M.}~\bibnamefont
			{Reiher}}, \bibinfo {author} {\bibfnamefont {N.}~\bibnamefont {Wiebe}},
		\bibinfo {author} {\bibfnamefont {K.~M.}\ \bibnamefont {Svore}}, \bibinfo
		{author} {\bibfnamefont {D.}~\bibnamefont {Wecker}}, \ and\ \bibinfo {author}
		{\bibfnamefont {M.}~\bibnamefont {Troyer}},\ }\href@noop {} {\bibfield
		{journal} {\bibinfo  {journal} {Proceedings of the National Academy of
				Sciences}\ }\textbf {\bibinfo {volume} {114}},\ \bibinfo {pages} {7555}
		(\bibinfo {year} {2017})}\BibitemShut {NoStop}%
	\bibitem [{\citenamefont {Ladd}\ \emph {et~al.}(2010)\citenamefont {Ladd},
		\citenamefont {Jelezko}, \citenamefont {Laflamme}, \citenamefont {Nakamura},
		\citenamefont {Monroe},\ and\ \citenamefont {O’Brien}}]{ladd2010quantum}%
	\BibitemOpen
	\bibfield  {author} {\bibinfo {author} {\bibfnamefont {T.~D.}\ \bibnamefont
			{Ladd}}, \bibinfo {author} {\bibfnamefont {F.}~\bibnamefont {Jelezko}},
		\bibinfo {author} {\bibfnamefont {R.}~\bibnamefont {Laflamme}}, \bibinfo
		{author} {\bibfnamefont {Y.}~\bibnamefont {Nakamura}}, \bibinfo {author}
		{\bibfnamefont {C.}~\bibnamefont {Monroe}}, \ and\ \bibinfo {author}
		{\bibfnamefont {J.~L.}\ \bibnamefont {O’Brien}},\ }\href@noop {} {\bibfield
		{journal} {\bibinfo  {journal} {Nature}\ }\textbf {\bibinfo {volume}
			{464}},\ \bibinfo {pages} {45} (\bibinfo {year} {2010})}\BibitemShut
	{NoStop}%
	\bibitem [{\citenamefont {Meunier}\ \emph {et~al.}(2011)\citenamefont
		{Meunier}, \citenamefont {Calado},\ and\ \citenamefont
		{Vandersypen}}]{meunier2011efficient}%
	\BibitemOpen
	\bibfield  {author} {\bibinfo {author} {\bibfnamefont {T.}~\bibnamefont
			{Meunier}}, \bibinfo {author} {\bibfnamefont {V.~E.}\ \bibnamefont {Calado}},
		\ and\ \bibinfo {author} {\bibfnamefont {L.~M.~K.}\ \bibnamefont
			{Vandersypen}},\ }\href@noop {} {\bibfield  {journal} {\bibinfo  {journal}
			{Physical Review B}\ }\textbf {\bibinfo {volume} {83}},\ \bibinfo {pages}
		{121403} (\bibinfo {year} {2011})}\BibitemShut {NoStop}%
	\bibitem [{\citenamefont {Kawakami}\ \emph {et~al.}(2014)\citenamefont
		{Kawakami}, \citenamefont {Scarlino}, \citenamefont {Ward}, \citenamefont
		{Braakman}, \citenamefont {Savage}, \citenamefont {Lagally}, \citenamefont
		{Friesen}, \citenamefont {Coppersmith}, \citenamefont {Eriksson},\ and\
		\citenamefont {Vandersypen}}]{kawakami_electrical_2014}%
	\BibitemOpen
	\bibfield  {author} {\bibinfo {author} {\bibfnamefont {E.}~\bibnamefont
			{Kawakami}}, \bibinfo {author} {\bibfnamefont {P.}~\bibnamefont {Scarlino}},
		\bibinfo {author} {\bibfnamefont {D.~R.}\ \bibnamefont {Ward}}, \bibinfo
		{author} {\bibfnamefont {F.~R.}\ \bibnamefont {Braakman}}, \bibinfo {author}
		{\bibfnamefont {D.~E.}\ \bibnamefont {Savage}}, \bibinfo {author}
		{\bibfnamefont {M.~G.}\ \bibnamefont {Lagally}}, \bibinfo {author}
		{\bibfnamefont {M.}~\bibnamefont {Friesen}}, \bibinfo {author} {\bibfnamefont
			{S.~N.}\ \bibnamefont {Coppersmith}}, \bibinfo {author} {\bibfnamefont
			{M.~A.}\ \bibnamefont {Eriksson}}, \ and\ \bibinfo {author} {\bibfnamefont
			{L.~M.~K.}\ \bibnamefont {Vandersypen}},\ }\href {\doibase
		10.1038/nnano.2014.153} {\bibfield  {journal} {\bibinfo  {journal} {Nature
				Nanotechnology}\ }\textbf {\bibinfo {volume} {9}},\ \bibinfo {pages} {666}
		(\bibinfo {year} {2014})}\BibitemShut {NoStop}%
	\bibitem [{\citenamefont {Yoneda}\ \emph {et~al.}(2018)\citenamefont {Yoneda},
		\citenamefont {Takeda}, \citenamefont {Otsuka}, \citenamefont {Nakajima},
		\citenamefont {Delbecq}, \citenamefont {Allison}, \citenamefont {Honda},
		\citenamefont {Kodera}, \citenamefont {Oda}, \citenamefont {Hoshi},
		\citenamefont {Usami}, \citenamefont {Itoh},\ and\ \citenamefont
		{Tarucha}}]{yoneda2018quantum}%
	\BibitemOpen
	\bibfield  {author} {\bibinfo {author} {\bibfnamefont {J.}~\bibnamefont
			{Yoneda}}, \bibinfo {author} {\bibfnamefont {K.}~\bibnamefont {Takeda}},
		\bibinfo {author} {\bibfnamefont {T.}~\bibnamefont {Otsuka}}, \bibinfo
		{author} {\bibfnamefont {T.}~\bibnamefont {Nakajima}}, \bibinfo {author}
		{\bibfnamefont {M.~R.}\ \bibnamefont {Delbecq}}, \bibinfo {author}
		{\bibfnamefont {G.}~\bibnamefont {Allison}}, \bibinfo {author} {\bibfnamefont
			{T.}~\bibnamefont {Honda}}, \bibinfo {author} {\bibfnamefont
			{T.}~\bibnamefont {Kodera}}, \bibinfo {author} {\bibfnamefont
			{S.}~\bibnamefont {Oda}}, \bibinfo {author} {\bibfnamefont {Y.}~\bibnamefont
			{Hoshi}}, \bibinfo {author} {\bibfnamefont {N.}~\bibnamefont {Usami}},
		\bibinfo {author} {\bibfnamefont {K.~M.}\ \bibnamefont {Itoh}}, \ and\
		\bibinfo {author} {\bibfnamefont {S.}~\bibnamefont {Tarucha}},\ }\href@noop
	{} {\bibfield  {journal} {\bibinfo  {journal} {Nature Nanotechnology}\
		}\textbf {\bibinfo {volume} {13}},\ \bibinfo {pages} {102} (\bibinfo {year}
		{2018})}\BibitemShut {NoStop}%
	\bibitem [{\citenamefont {Sigillito}\ \emph {et~al.}(2019)\citenamefont
		{Sigillito}, \citenamefont {Gullans}, \citenamefont {Edge}, \citenamefont
		{Borselli},\ and\ \citenamefont {Petta}}]{sigillito2019coherent}%
	\BibitemOpen
	\bibfield  {author} {\bibinfo {author} {\bibfnamefont {A.~J.}\ \bibnamefont
			{Sigillito}}, \bibinfo {author} {\bibfnamefont {M.~J.}\ \bibnamefont
			{Gullans}}, \bibinfo {author} {\bibfnamefont {L.~F.}\ \bibnamefont {Edge}},
		\bibinfo {author} {\bibfnamefont {M.}~\bibnamefont {Borselli}}, \ and\
		\bibinfo {author} {\bibfnamefont {J.~R.}\ \bibnamefont {Petta}},\ }\href@noop
	{} {\bibfield  {journal} {\bibinfo  {journal} {npj Quantum Information}\
		}\textbf {\bibinfo {volume} {5}},\ \bibinfo {pages} {1} (\bibinfo {year}
		{2019})}\BibitemShut {NoStop}%
	\bibitem [{\citenamefont {Barenco}\ \emph {et~al.}(1995)\citenamefont
		{Barenco}, \citenamefont {Bennett}, \citenamefont {Cleve}, \citenamefont
		{DiVincenzo}, \citenamefont {Margolus}, \citenamefont {Shor}, \citenamefont
		{Sleator}, \citenamefont {Smolin},\ and\ \citenamefont
		{Weinfurter}}]{barenco1995elementary}%
	\BibitemOpen
	\bibfield  {author} {\bibinfo {author} {\bibfnamefont {A.}~\bibnamefont
			{Barenco}}, \bibinfo {author} {\bibfnamefont {C.~H.}\ \bibnamefont
			{Bennett}}, \bibinfo {author} {\bibfnamefont {R.}~\bibnamefont {Cleve}},
		\bibinfo {author} {\bibfnamefont {D.~P.}\ \bibnamefont {DiVincenzo}},
		\bibinfo {author} {\bibfnamefont {N.}~\bibnamefont {Margolus}}, \bibinfo
		{author} {\bibfnamefont {P.}~\bibnamefont {Shor}}, \bibinfo {author}
		{\bibfnamefont {T.}~\bibnamefont {Sleator}}, \bibinfo {author} {\bibfnamefont
			{J.~A.}\ \bibnamefont {Smolin}}, \ and\ \bibinfo {author} {\bibfnamefont
			{H.}~\bibnamefont {Weinfurter}},\ }\href@noop {} {\bibfield  {journal}
		{\bibinfo  {journal} {Physical Review A}\ }\textbf {\bibinfo {volume} {52}},\
		\bibinfo {pages} {3457} (\bibinfo {year} {1995})}\BibitemShut {NoStop}%
	\bibitem [{\citenamefont {Veldhorst}\ \emph {et~al.}(2017)\citenamefont
		{Veldhorst}, \citenamefont {Eenink}, \citenamefont {Yang},\ and\
		\citenamefont {Dzurak}}]{veldhorst_silicon_2017}%
	\BibitemOpen
	\bibfield  {author} {\bibinfo {author} {\bibfnamefont {M.}~\bibnamefont
			{Veldhorst}}, \bibinfo {author} {\bibfnamefont {H.~G.~J.}\ \bibnamefont
			{Eenink}}, \bibinfo {author} {\bibfnamefont {C.~H.}\ \bibnamefont {Yang}}, \
		and\ \bibinfo {author} {\bibfnamefont {A.~S.}\ \bibnamefont {Dzurak}},\
	}\href {\doibase 10.1038/s41467-017-01905-6} {\bibfield  {journal} {\bibinfo
			{journal} {Nature Communication}\ }\textbf {\bibinfo {volume} {8}},\ \bibinfo
		{pages} {1766} (\bibinfo {year} {2017})}\BibitemShut {NoStop}%
	\bibitem [{\citenamefont {Li}\ \emph {et~al.}(2018)\citenamefont {Li},
		\citenamefont {Petit}, \citenamefont {Franke}, \citenamefont {Dehollain},
		\citenamefont {Helsen}, \citenamefont {Steudtner}, \citenamefont {Thomas},
		\citenamefont {Yoscovits}, \citenamefont {Singh}, \citenamefont {Wehner},
		\citenamefont {Vandersypen}, \citenamefont {Clarke},\ and\ \citenamefont
		{Veldhorst}}]{li2018crossbar}%
	\BibitemOpen
	\bibfield  {author} {\bibinfo {author} {\bibfnamefont {R.}~\bibnamefont
			{Li}}, \bibinfo {author} {\bibfnamefont {L.}~\bibnamefont {Petit}}, \bibinfo
		{author} {\bibfnamefont {D.~P.}\ \bibnamefont {Franke}}, \bibinfo {author}
		{\bibfnamefont {J.~P.}\ \bibnamefont {Dehollain}}, \bibinfo {author}
		{\bibfnamefont {J.}~\bibnamefont {Helsen}}, \bibinfo {author} {\bibfnamefont
			{M.}~\bibnamefont {Steudtner}}, \bibinfo {author} {\bibfnamefont {N.~K.}\
			\bibnamefont {Thomas}}, \bibinfo {author} {\bibfnamefont {Z.~R.}\
			\bibnamefont {Yoscovits}}, \bibinfo {author} {\bibfnamefont {K.~J.}\
			\bibnamefont {Singh}}, \bibinfo {author} {\bibfnamefont {S.}~\bibnamefont
			{Wehner}}, \bibinfo {author} {\bibfnamefont {L.~M.~K.}\ \bibnamefont
			{Vandersypen}}, \bibinfo {author} {\bibfnamefont {J.~S.}\ \bibnamefont
			{Clarke}}, \ and\ \bibinfo {author} {\bibfnamefont {M.}~\bibnamefont
			{Veldhorst}},\ }\href@noop {} {\bibfield  {journal} {\bibinfo  {journal}
			{Science Advances}\ }\textbf {\bibinfo {volume} {4}},\ \bibinfo {pages}
		{eaar3960} (\bibinfo {year} {2018})}\BibitemShut {NoStop}%
	\bibitem [{\citenamefont {Lawrie}\ \emph {et~al.}(2020)\citenamefont {Lawrie},
		\citenamefont {Eenink}, \citenamefont {Hendrickx}, \citenamefont {Boter},
		\citenamefont {Petit}, \citenamefont {Amitonov}, \citenamefont {Lodari},
		\citenamefont {Paquelet~Wuetz}, \citenamefont {Volk}, \citenamefont
		{Philips}, \citenamefont {Droulers}, \citenamefont {Kalhor}, \citenamefont
		{van Riggelen}, \citenamefont {Brousse}, \citenamefont {Sammak},
		\citenamefont {Vandersypen}, \citenamefont {Scappucci},\ and\ \citenamefont
		{Veldhorst}}]{lawrie2020quantum}%
	\BibitemOpen
	\bibfield  {author} {\bibinfo {author} {\bibfnamefont {W.~I.~L.}\
			\bibnamefont {Lawrie}}, \bibinfo {author} {\bibfnamefont {H.~G.~J.}\
			\bibnamefont {Eenink}}, \bibinfo {author} {\bibfnamefont {N.~W.}\
			\bibnamefont {Hendrickx}}, \bibinfo {author} {\bibfnamefont {J.~M.}\
			\bibnamefont {Boter}}, \bibinfo {author} {\bibfnamefont {L.}~\bibnamefont
			{Petit}}, \bibinfo {author} {\bibfnamefont {S.~V.}\ \bibnamefont {Amitonov}},
		\bibinfo {author} {\bibfnamefont {M.}~\bibnamefont {Lodari}}, \bibinfo
		{author} {\bibfnamefont {B.}~\bibnamefont {Paquelet~Wuetz}}, \bibinfo
		{author} {\bibfnamefont {C.}~\bibnamefont {Volk}}, \bibinfo {author}
		{\bibfnamefont {S.~G.~J.}\ \bibnamefont {Philips}}, \bibinfo {author}
		{\bibfnamefont {G.}~\bibnamefont {Droulers}}, \bibinfo {author}
		{\bibfnamefont {N.}~\bibnamefont {Kalhor}}, \bibinfo {author} {\bibfnamefont
			{F.}~\bibnamefont {van Riggelen}}, \bibinfo {author} {\bibfnamefont
			{D.}~\bibnamefont {Brousse}}, \bibinfo {author} {\bibfnamefont
			{A.}~\bibnamefont {Sammak}}, \bibinfo {author} {\bibfnamefont {L.~M.~K.}\
			\bibnamefont {Vandersypen}}, \bibinfo {author} {\bibfnamefont
			{G.}~\bibnamefont {Scappucci}}, \ and\ \bibinfo {author} {\bibfnamefont
			{M.}~\bibnamefont {Veldhorst}},\ }\href@noop {} {\bibfield  {journal}
		{\bibinfo  {journal} {Applied Physics Letters}\ }\textbf {\bibinfo {volume}
			{116}},\ \bibinfo {pages} {080501} (\bibinfo {year} {2020})}\BibitemShut
	{NoStop}%
	\bibitem [{\citenamefont {Urdampilleta}\ \emph {et~al.}(2019)\citenamefont
		{Urdampilleta}, \citenamefont {Niegemann}, \citenamefont {Chanrion},
		\citenamefont {Jadot}, \citenamefont {Spence}, \citenamefont {Mortemousque},
		\citenamefont {Hutin}, \citenamefont {Bertrand}, \citenamefont {Barraud},
		\citenamefont {Maurand}, \citenamefont {Sanquer}, \citenamefont {Jehl},
		\citenamefont {De~Franceschi}, \citenamefont {Vinet},\ and\ \citenamefont
		{Meunier}}]{urdampilleta2018gate}%
	\BibitemOpen
	\bibfield  {author} {\bibinfo {author} {\bibfnamefont {M.}~\bibnamefont
			{Urdampilleta}}, \bibinfo {author} {\bibfnamefont {D.~J.}\ \bibnamefont
			{Niegemann}}, \bibinfo {author} {\bibfnamefont {E.}~\bibnamefont {Chanrion}},
		\bibinfo {author} {\bibfnamefont {B.}~\bibnamefont {Jadot}}, \bibinfo
		{author} {\bibfnamefont {C.}~\bibnamefont {Spence}}, \bibinfo {author}
		{\bibfnamefont {P.}~\bibnamefont {Mortemousque}}, \bibinfo {author}
		{\bibfnamefont {L.}~\bibnamefont {Hutin}}, \bibinfo {author} {\bibfnamefont
			{B.}~\bibnamefont {Bertrand}}, \bibinfo {author} {\bibfnamefont
			{S.}~\bibnamefont {Barraud}}, \bibinfo {author} {\bibfnamefont
			{R.}~\bibnamefont {Maurand}}, \bibinfo {author} {\bibfnamefont
			{M.}~\bibnamefont {Sanquer}}, \bibinfo {author} {\bibfnamefont
			{X.}~\bibnamefont {Jehl}}, \bibinfo {author} {\bibfnamefont {S.~D.}\
			\bibnamefont {De~Franceschi}}, \bibinfo {author} {\bibfnamefont
			{M.}~\bibnamefont {Vinet}}, \ and\ \bibinfo {author} {\bibfnamefont
			{T.}~\bibnamefont {Meunier}},\ }\href@noop {} {\bibfield  {journal} {\bibinfo
			{journal} {Nature Nanotechnology}\ }\textbf {\bibinfo {volume} {14}},\
		\bibinfo {pages} {737} (\bibinfo {year} {2019})}\BibitemShut {NoStop}%
	\bibitem [{\citenamefont {Russ}\ \emph {et~al.}(2018)\citenamefont {Russ},
		\citenamefont {Zajac}, \citenamefont {Sigillito}, \citenamefont {Borjans},
		\citenamefont {Taylor}, \citenamefont {Petta},\ and\ \citenamefont
		{Burkard}}]{russ2018high}%
	\BibitemOpen
	\bibfield  {author} {\bibinfo {author} {\bibfnamefont {M.}~\bibnamefont
			{Russ}}, \bibinfo {author} {\bibfnamefont {D.~M.}\ \bibnamefont {Zajac}},
		\bibinfo {author} {\bibfnamefont {A.~J.}\ \bibnamefont {Sigillito}}, \bibinfo
		{author} {\bibfnamefont {F.}~\bibnamefont {Borjans}}, \bibinfo {author}
		{\bibfnamefont {J.~M.}\ \bibnamefont {Taylor}}, \bibinfo {author}
		{\bibfnamefont {J.~R.}\ \bibnamefont {Petta}}, \ and\ \bibinfo {author}
		{\bibfnamefont {G.}~\bibnamefont {Burkard}},\ }\href@noop {} {\bibfield
		{journal} {\bibinfo  {journal} {Physical Review B}\ }\textbf {\bibinfo
			{volume} {97}},\ \bibinfo {pages} {085421} (\bibinfo {year}
		{2018})}\BibitemShut {NoStop}%
	\bibitem [{\citenamefont {Burkard}\ \emph {et~al.}(1999)\citenamefont
		{Burkard}, \citenamefont {Loss}, \citenamefont {DiVincenzo},\ and\
		\citenamefont {Smolin}}]{burkard1999physical}%
	\BibitemOpen
	\bibfield  {author} {\bibinfo {author} {\bibfnamefont {G.}~\bibnamefont
			{Burkard}}, \bibinfo {author} {\bibfnamefont {D.}~\bibnamefont {Loss}},
		\bibinfo {author} {\bibfnamefont {D.~P.}\ \bibnamefont {DiVincenzo}}, \ and\
		\bibinfo {author} {\bibfnamefont {J.~A.}\ \bibnamefont {Smolin}},\
	}\href@noop {} {\bibfield  {journal} {\bibinfo  {journal} {Physical Review
				B}\ }\textbf {\bibinfo {volume} {60}},\ \bibinfo {pages} {11404} (\bibinfo
		{year} {1999})}\BibitemShut {NoStop}%
	\bibitem [{\citenamefont {Maune}\ \emph {et~al.}(2012)\citenamefont {Maune},
		\citenamefont {Borselli}, \citenamefont {Huang}, \citenamefont {Ladd},
		\citenamefont {Deelman}, \citenamefont {Holabird}, \citenamefont {Kiselev},
		\citenamefont {Alvarado-Rodriguez}, \citenamefont {Ross}, \citenamefont
		{Schmitz}, \citenamefont {Sokolich}, \citenamefont {Watson}, \citenamefont
		{Gyure},\ and\ \citenamefont {Hunter}}]{maune2012coherent}%
	\BibitemOpen
	\bibfield  {author} {\bibinfo {author} {\bibfnamefont {B.~M.}\ \bibnamefont
			{Maune}}, \bibinfo {author} {\bibfnamefont {M.~G.}\ \bibnamefont {Borselli}},
		\bibinfo {author} {\bibfnamefont {B.}~\bibnamefont {Huang}}, \bibinfo
		{author} {\bibfnamefont {T.~D.}\ \bibnamefont {Ladd}}, \bibinfo {author}
		{\bibfnamefont {P.~W.}\ \bibnamefont {Deelman}}, \bibinfo {author}
		{\bibfnamefont {K.~S.}\ \bibnamefont {Holabird}}, \bibinfo {author}
		{\bibfnamefont {A.~A.}\ \bibnamefont {Kiselev}}, \bibinfo {author}
		{\bibfnamefont {I.}~\bibnamefont {Alvarado-Rodriguez}}, \bibinfo {author}
		{\bibfnamefont {R.~S.}\ \bibnamefont {Ross}}, \bibinfo {author}
		{\bibfnamefont {A.~E.}\ \bibnamefont {Schmitz}}, \bibinfo {author}
		{\bibfnamefont {M.}~\bibnamefont {Sokolich}}, \bibinfo {author}
		{\bibfnamefont {C.}~\bibnamefont {Watson}}, \bibinfo {author} {\bibfnamefont
			{M.}~\bibnamefont {Gyure}}, \ and\ \bibinfo {author} {\bibfnamefont
			{A.}~\bibnamefont {Hunter}},\ }\href@noop {} {\bibfield  {journal} {\bibinfo
			{journal} {Nature}\ }\textbf {\bibinfo {volume} {481}},\ \bibinfo {pages}
		{344} (\bibinfo {year} {2012})}\BibitemShut {NoStop}%
	\bibitem [{\citenamefont {He}\ \emph {et~al.}(2019)\citenamefont {He},
		\citenamefont {Gorman}, \citenamefont {Keith}, \citenamefont {Kranz},
		\citenamefont {Keizer},\ and\ \citenamefont {Simmons}}]{he2019two}%
	\BibitemOpen
	\bibfield  {author} {\bibinfo {author} {\bibfnamefont {Y.}~\bibnamefont
			{He}}, \bibinfo {author} {\bibfnamefont {S.~K.}\ \bibnamefont {Gorman}},
		\bibinfo {author} {\bibfnamefont {D.}~\bibnamefont {Keith}}, \bibinfo
		{author} {\bibfnamefont {L.}~\bibnamefont {Kranz}}, \bibinfo {author}
		{\bibfnamefont {J.~G.}\ \bibnamefont {Keizer}}, \ and\ \bibinfo {author}
		{\bibfnamefont {M.~Y.}\ \bibnamefont {Simmons}},\ }\href@noop {} {\bibfield
		{journal} {\bibinfo  {journal} {Nature}\ }\textbf {\bibinfo {volume} {571}},\
		\bibinfo {pages} {371} (\bibinfo {year} {2019})}\BibitemShut {NoStop}%
	\bibitem [{\citenamefont {Martinis}\ and\ \citenamefont
		{Geller}(2014)}]{martinis2014fast}%
	\BibitemOpen
	\bibfield  {author} {\bibinfo {author} {\bibfnamefont {J.~M.}\ \bibnamefont
			{Martinis}}\ and\ \bibinfo {author} {\bibfnamefont {M.~R.}\ \bibnamefont
			{Geller}},\ }\href@noop {} {\bibfield  {journal} {\bibinfo  {journal}
			{Physical Review A}\ }\textbf {\bibinfo {volume} {90}},\ \bibinfo {pages}
		{022307} (\bibinfo {year} {2014})}\BibitemShut {NoStop}%
	\bibitem [{\citenamefont {G{\"u}ng{\"o}rd{\"u}}\ and\ \citenamefont
		{Kestner}(2018)}]{gungordu2018pulse}%
	\BibitemOpen
	\bibfield  {author} {\bibinfo {author} {\bibfnamefont {U.}~\bibnamefont
			{G{\"u}ng{\"o}rd{\"u}}}\ and\ \bibinfo {author} {\bibfnamefont
			{J.}~\bibnamefont {Kestner}},\ }\href@noop {} {\bibfield  {journal} {\bibinfo
			{journal} {Physical Review B}\ }\textbf {\bibinfo {volume} {98}},\ \bibinfo
		{pages} {165301} (\bibinfo {year} {2018})}\BibitemShut {NoStop}%
	\bibitem [{\citenamefont {Calderon-Vargas}\ \emph {et~al.}(2019)\citenamefont
		{Calderon-Vargas}, \citenamefont {Barron}, \citenamefont {Deng},
		\citenamefont {Sigillito}, \citenamefont {Barnes},\ and\ \citenamefont
		{Economou}}]{calderon2019fast}%
	\BibitemOpen
	\bibfield  {author} {\bibinfo {author} {\bibfnamefont {F.}~\bibnamefont
			{Calderon-Vargas}}, \bibinfo {author} {\bibfnamefont {G.~S.}\ \bibnamefont
			{Barron}}, \bibinfo {author} {\bibfnamefont {X.-H.}\ \bibnamefont {Deng}},
		\bibinfo {author} {\bibfnamefont {A.}~\bibnamefont {Sigillito}}, \bibinfo
		{author} {\bibfnamefont {E.}~\bibnamefont {Barnes}}, \ and\ \bibinfo {author}
		{\bibfnamefont {S.~E.}\ \bibnamefont {Economou}},\ }\href@noop {} {\bibfield
		{journal} {\bibinfo  {journal} {Physical Review B}\ }\textbf {\bibinfo
			{volume} {100}},\ \bibinfo {pages} {035304} (\bibinfo {year}
		{2019})}\BibitemShut {NoStop}%
	\bibitem [{\citenamefont {G{\"u}ng{\"o}rd{\"u}}\ and\ \citenamefont
		{Kestner}(2019)}]{gungordu2019analytically}%
	\BibitemOpen
	\bibfield  {author} {\bibinfo {author} {\bibfnamefont {U.}~\bibnamefont
			{G{\"u}ng{\"o}rd{\"u}}}\ and\ \bibinfo {author} {\bibfnamefont
			{J.}~\bibnamefont {Kestner}},\ }\href@noop {} {\bibfield  {journal} {\bibinfo
			{journal} {Physical Review A}\ }\textbf {\bibinfo {volume} {100}},\ \bibinfo
		{pages} {062310} (\bibinfo {year} {2019})}\BibitemShut {NoStop}%
	\bibitem [{\citenamefont {G{\"u}ng{\"o}rd{\"u}}\ and\ \citenamefont
		{Kestner}(2020)}]{gungordu2020robust}%
	\BibitemOpen
	\bibfield  {author} {\bibinfo {author} {\bibfnamefont {U.}~\bibnamefont
			{G{\"u}ng{\"o}rd{\"u}}}\ and\ \bibinfo {author} {\bibfnamefont
			{J.}~\bibnamefont {Kestner}},\ }\href@noop {} {\bibfield  {journal} {\bibinfo
			{journal} {Physical Review B}\ }\textbf {\bibinfo {volume} {101}},\ \bibinfo
		{pages} {155301} (\bibinfo {year} {2020})}\BibitemShut {NoStop}%
	\bibitem [{\citenamefont {Vandersypen}\ and\ \citenamefont
		{Chuang}(2005)}]{vandersypen2005nmr}%
	\BibitemOpen
	\bibfield  {author} {\bibinfo {author} {\bibfnamefont {L.~M.~K.}\
			\bibnamefont {Vandersypen}}\ and\ \bibinfo {author} {\bibfnamefont {I.~L.}\
			\bibnamefont {Chuang}},\ }\href@noop {} {\bibfield  {journal} {\bibinfo
			{journal} {Reviews of Modern Physics}\ }\textbf {\bibinfo {volume} {76}},\
		\bibinfo {pages} {1037} (\bibinfo {year} {2005})}\BibitemShut {NoStop}%
\end{thebibliography}
\end{document}